\documentclass[conference]{IEEEtran}
\IEEEoverridecommandlockouts

\usepackage{cite}
\usepackage{amsmath,amssymb,amsfonts,booktabs,multirow}
\usepackage{algorithmic}
\usepackage{graphicx}
\usepackage{textcomp}
\usepackage{xcolor}
\usepackage{lineno,hyperref}
\usepackage[obeyspaces]{xurl}
\usepackage{comment}
\def\BibTeX{{\rm B\kern-.05em{\sc i\kern-.025em b}\kern-.08em
    T\kern-.1667em\lower.7ex\hbox{E}\kern-.125emX}}
\begin{document}

\title{A Universal Harmonic Discriminator for High-quality GAN-based Vocoder}

\author{\IEEEauthorblockN{Nan Xu}
\IEEEauthorblockA{\textit{Alibaba Digital Media \&} \\
\textit{Entertainment Group}\\
Beijing, China} 
\and
\IEEEauthorblockN{Zhaolong Huang}
\IEEEauthorblockA{\textit{Alibaba Digital Media \&} \\
\textit{Entertainment Group}\\
Beijing, China}
\and
\IEEEauthorblockN{Xiao Zeng*} \thanks{*Corresponding author. Email: zengxiao@alibaba-inc.com}
\IEEEauthorblockA{\textit{Alibaba Digital Media \&} \\
\textit{Entertainment Group}\\
Beijing, China}
}

\maketitle

\begin{abstract}
\sloppy
With the emergence of GAN-based vocoders, the discriminator, as a crucial component, has been developed recently. In our work, we focus on improving the time-frequency based discriminator. Particularly, Short-Time Fourier Transform (STFT) representation is usually used as input of time-frequency based discriminator. However, the STFT spectrogram has the same frequency resolution at different frequency bins, which results in an inferior performance, especially for singing voices. Motivated by this, we propose a universal harmonic discriminator for dynamic frequency resolution modeling and harmonic tracking. Specifically, we design a harmonic filter with learnable triangular band-pass filter banks, where each frequency bin has a flexible bandwidth. Additionally, we add a half-harmonic to capture fine-grained harmonic relationships at low-frequency band. Experiments on speech and singing datasets validate the effectiveness of the proposed discriminator on both subjective and objective metrics.
\end{abstract}


\section{Introduction}
\label{int}

\sloppy
With the advancements in large language model, the performance of the multimodal interaction has been significantly improved in recent years. Speech, as one of the interaction entrances, plays a crucial role in the multimodal domain \cite{lee-etal-2022-direct}. Therefore, how to generate speech or audio response has become a hot research field. Text-to-speech (TTS) \cite{ren2019fastspeech,bai20223,jiang2024mega} or text-to-audio (TTA) \cite{huang2023make,liu2024audioldm} is thus an underlying task, which targets to generate speech from the corresponding input text. These technologies initially produce intermediate acoustic features and then convert these features into the speech waveforms by utilizing the vocoders. Therefore, high-quality vocoders are crucial for the expressive speech generation. Recent advancements in deep learning bring the surprising improvements for neural vocoders in terms of generating high perceptual quality and intelligibility speech waveforms \cite{Lee2025period,kong2020hifi,leebigvgan}. Among these methods, GAN-based methods, balancing the synthesis quality and inference speed, are widely utilized in vocoder tasks.

\sloppy
The generator and discriminator play the crucial roles in the adversarial training process of GANs. The former is predominantly driven by two kinds of types: direct generation methods \cite{kumar2019melgan,kong2020hifi,tian2020tfgan,Kazuki2021hwg,jang2021univnet,leebigvgan,song2023dspgan,li2023snakegan,bak2023avocodo} and inverse short-time Fourier transform (iSTFT) based methods \cite{ai2020neural,kaneko2022istftnet,kaneko2023istftnet2,ai2023apnet,du2023apnet2,siuzdak2023vocos}. Specifically, the direct generation methods usually employ convolution neural network (CNN) architectures with temporal transposed convolution layers to directly sequential upsample the melspectrogram representation to the raw waveform. The iSTFT-based methods usually predict the high-dimensional full-band Fourier spectral coefficients, i.e., magnitude and phase spectrums, and next apply iSTFT to generate the high-quality waveform. In addition, the discriminators mainly contain the time and time-frequency representation based methods. The multi-period discriminator (MPD) \cite{kong2020hifi} with the periodic folding and the multi-scale discriminator (MSD) \cite{kumar2019melgan} with the averaging pooling are the most typical time representation based discriminators. In contrast, time-frequency representation based methods operate on the frequency features. TFGAN \cite{tian2020tfgan} employs a STFT module followed by a frequency discriminator. Harmonic WaveGAN \cite{Kazuki2021hwg} uses the harmonic convolution to learn the harmonic structure, which increases the computation time during the training process. The UnivNet \cite{jang2021univnet} and Vocos \cite{siuzdak2023vocos} utilize the multi-resolution spectrogram discriminator on multiple equant frequency bands. Encodec \cite{defossez2022high} verifies the effectiveness of the Multi-Scale STFT (MS-STFT) discriminator with different window lengths. Avocodo \cite{bak2023avocodo} designs a collaborative multi-scale multi-band discriminator and a sub-band discriminator with the pseudo quadrature mirror filter (PQMF) \cite{nguyen1994near}. The Constant-Q Transform (CQT) \cite{brown1992efficient} representation is also used as input of the discriminator, contributing to the dynamic frequency resolution for different frequency bands \cite{gu2024multi}.

\sloppy
In this paper, we mainly focus on the improvement of the time-frequency based discriminator. The aforementioned time-frequency based discriminators are predominantly driven by STFT features. However, the STFT representation usually utilizes equally spaced filters to extract spectrograms, which leads to the same frequency resolution across different frequency bins and further limits the performance improvement of vocoders. For example, expressive singing voices usually require flexible attention for different harmonics. The fixed frequency resolution cannot achieve the accurate reconstruction of these harmonics, leading to the suboptimal performance. While CQT representation has the dynamic frequency resolution, it suffers from the difficulty of odd harmonics modeling and temporally asynchronous problem \cite{brown1992efficient,schorkhuber2010constant}. Gu et al. \cite{gu2024multi} uses the sub-band processing module (SBP) to learn temporally synchronized representations. However, each octave requires an SBP for both real and imaginary part features, which reduces the scalability of the discriminator architecture. Additionally, odd harmonics are still difficult to capture in the CQT spectrogram.

\sloppy
To address these limitations, we propose a {\bf Univ}ersal {\bf H}armonic {\bf D}iscriminator for GAN-based vocoders, termed {\bf UnivHD}. Specifically, we remain the STFT operator followed by a designed harmonic filter with the learnable band-pass filter banks. For this discriminator, rather than using the same resolution in the frequency domain, the proposed UnivHD owns the dynamic frequency resolution at different frequency bins. In other words, the low-frequency band can achieve a higher frequency resolution, contributing to the pitch modeling. At the high-frequency band, bandwidths become wider, i.e., a higher time resolution, which can achieve the significant advantages for better fast-changing harmonic tracking. In addition, by taking the STFT with fixed window length as input, the temporally asynchronous problem in the CQT spectrogram is avoided. Moreover, the center frequencies of this harmonic filter are scaled by the harmonic orders, which can successfully learn odd harmonics in the complete harmonic space. A half-harmonic representation is also added to learn the energy at low-frequency band. Furthermore, for the discriminator architecture, we utilize the combination of the depthwise separable convolution and normal convolution to model the intra-harmonic and inter-harmonic information in speech, respectively. 

\sloppy
The primary contributions in this paper are as follows:
\begin{itemize}
\item We propose the UnivHD, trained with the complete harmonic space that has the dynamic frequency resolution. Each harmonic filter preserves the primary harmonic structure and contributes to better synthesis quality of expressive speech. 
\item We design an effective network architecture, where the depthwise separable convolution and normal convolution are combined to simultaneously capture the relationships of intra-harmonic and inter-harmonic.
\item We conduct the extensive experiments and results demonstrate that the UnivHD achieves the competitive speech quality in terms of both subjective and objective metrics. Furthermore, we also validate the effectiveness of the half-harmonic representation.
\end{itemize}

\sloppy
The rest parts of the proposed paper are organized as follows: In Section \ref{work}, the related general filter frameworks are introduced. Next in Section \ref{format}, we will introduce the proposed harmonic discriminator method, including the learnable harmonic filter and discriminator network architecture. Furthermore, experiment results are reported in Section \ref{exp}. Finally, Section \ref{con} is the conclusion of this paper.

\begin{figure}[t]
\centering
\includegraphics[width=0.45\textwidth]{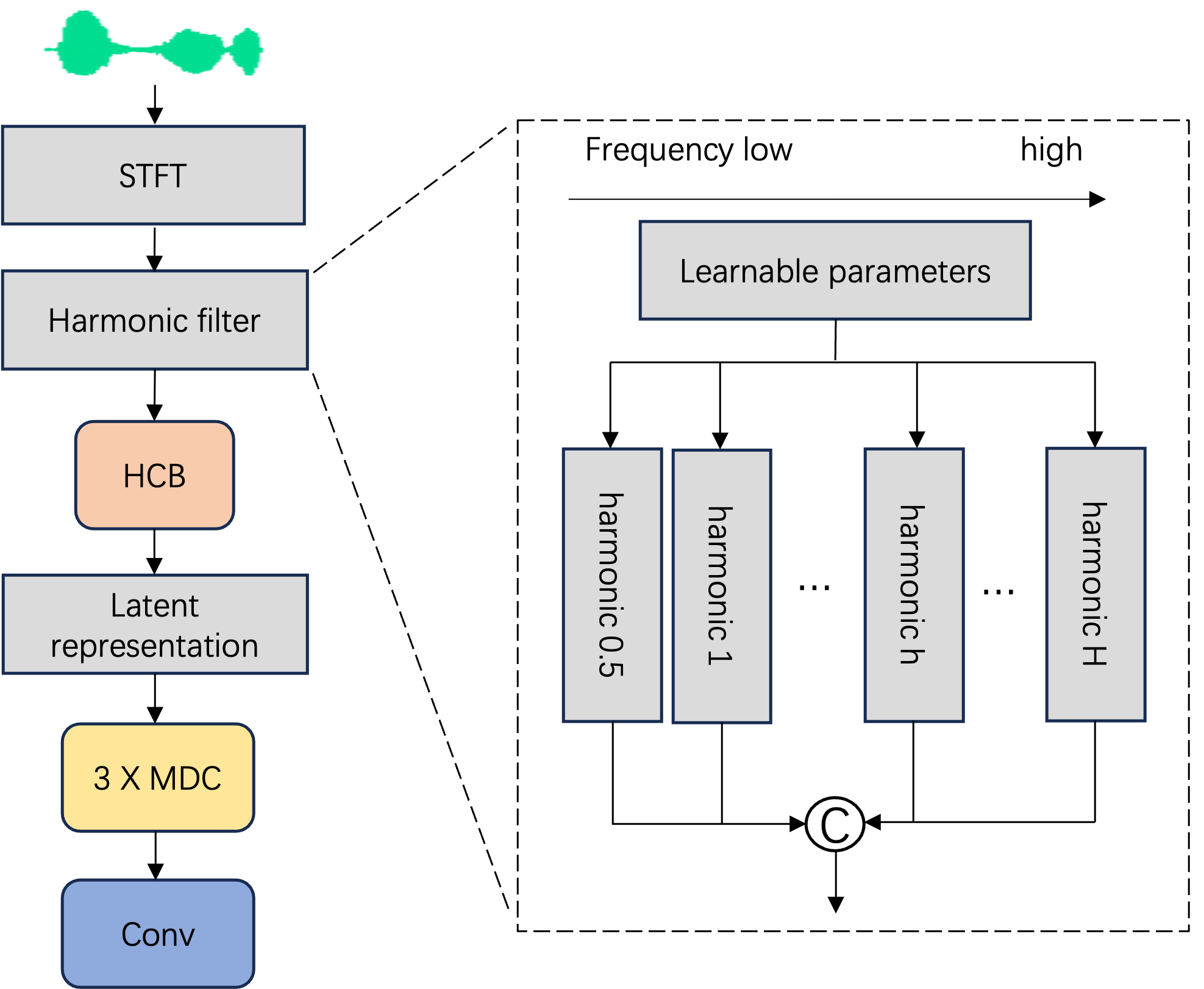}
\caption{The overall architecture of the universal harmonic discriminator. "har" denotes the proposed harmonic filter. HCB denotes the hybrid convolution block and MDC is the multi-scale dilated convolution. “C” refers to the concatenation operator. The harmonic filter takes STFT representation as input and has the learnable bandwidth parameters.}
\label{mhd}
\end{figure}

\section{Related Work}
\label{work}
\sloppy
Most filters and transforms are applied and developed in speech signal processing \cite{brown1992efficient,nguyen1994near, pariente2020filterbank,won2020data,wang2003modified}. In this section, we will introduce the related STFT and CQT frameworks and explain their differences.

\sloppy
In general, following \cite{brown1992efficient,pariente2020filterbank}, the time-domain signal $ x $ can be transformed into frequency domain by convolving an analysis filter bank $ u_k\left(n\right) $, which can be written as follows: 
\begin{equation}
\label{eq1}
\setlength{\abovedisplayskip}{6pt}
\setlength{\belowdisplayskip}{6pt}
X\left(k,n\right)=\sum_{n=0}^{N-1}{x\left(n\right)\cdot{u_k}\left(n\right)},
\end{equation}
where $ N $ is the length of the analysis filter. For the STFT, 
\begin{equation}
\label{eq2}
\setlength{\abovedisplayskip}{6pt}
\setlength{\belowdisplayskip}{6pt}
{u_k}\left(n\right)={w_k}\left(n\right)e^{-j\frac{{2\pi}k}{N}n},
\end{equation}
where $ w_k\left(n\right) $ is the analysis  window.

In the case of the CQT representation, the analysis filter bank can be defined as follows:
\begin{equation}
\label{eq3}
\setlength{\abovedisplayskip}{6pt}
\setlength{\belowdisplayskip}{6pt}
u_k\left(n\right)=\frac{1}{N_k}w_k\left(n\right)e^{-j\frac{{2\pi}{Q_k}}{N_k}n},
\end{equation}
where $ N_k $ is the window length of the $ k $-th frequency bin and $ Q_k $ is the constant Q-factor:
\begin{equation}
\label{eq4}
\setlength{\abovedisplayskip}{6pt}
\setlength{\belowdisplayskip}{6pt}
Q_k={f_c^k}/{f_{bw}^k},
\end{equation}
where $ {f_{bw}^k} $ is the bandwidth and $ f_c^k $ is the center frequency that is defined as:
\begin{equation}
\label{eq5}
\setlength{\abovedisplayskip}{6pt}
\setlength{\belowdisplayskip}{6pt}
f_c^k=f_{min}\cdot2^\frac{k}{B},
\end{equation}
where the lowest frequency $ f_{min}=32.7 $ Hz (C1), $ k $ is the frequency index of the filter and $ B $ is the number of bins per octave.

\sloppy
Notably, for the STFT representation, the bandwidth is constant, resulting in the same frequency resolution at different frequency bins. Instead, due to the fixed $ Q_k $, the bandwidth of CQT is changed with the center frequency according to \eqref{eq4}, which means the frequency resolution is dynamic. However, due to the unfixed $ N_k $, the analysis filter is not temporally synchronized at different frequency bins, and the modules followed CQT representation need to be elaborately designed \cite{gu2024multi}. In addition, since the center frequency at the $ k $-th frequency bin is measured by using the \eqref{eq5}, the $ h $-th harmonic $ h\cdot{f_c^k} $ can be learned only when $ h=2^d $ ($ d $ is integer), which results in the difficulty of odd harmonics modeling. In this work, we propose a novel harmonic discriminator to address these problems.

\section{Method}
\label{format}
\sloppy
As illustrated in Fig.~\ref{mhd}, the UnivHD is built on the harmonic filter bank. These representations align the harmonically related contents that are learned by the followed network architecture. In this section, we will first introduce the proposed universal harmonic discriminator in Section \ref{ana},  and then in Section \ref{str}, we will describe the designed model structure.

\subsection{Universal Harmonic Discriminator}
\label{ana}
\sloppy
Most filters and transforms own the fixed filter bank with the invariant frequency resolution, which results in the insufficient optimization for the pitch and harmonic properties \cite{pariente2020filterbank,won2020data,wang2003modified,ho2024neural}. In contrast, the ideal filter should correctly learn the pitch and effectively track the harmonic changing. In other words, for the pitch property, a positive value can reflect the perceptible loudness at the fundamental frequency, and the zero value is given where there is no perceived pitch. In terms of the harmonic property, the input representation must align the harmonically related content that contributes to better tracking for fast-changing harmonics.

\sloppy
To address these problems, we design a novel harmonic discriminator with learnable band-pass filter $ \nabla_h $ which can be written as follows:
\begin{equation}
\label{eq9}
\setlength{\abovedisplayskip}{6pt}
\setlength{\belowdisplayskip}{6pt}
\nabla_h\left(f;f_c;f_{bw}^h\right)=\left[1-\frac{2\left|f-{h\cdot} f_c\right|}{f_{bw}^h}\right]_+,
\end{equation}
where $ \nabla_h $ is the $ h $-th harmonic filter with the triangular band-pass filter and its center frequency is $ h\cdot f_c $. The STFT representation in Section \ref{work} is used as the input of $ \nabla_h $ to obtain the harmonic tensor. $ \left[.\right]_+ $ refers to the rectified linear function. In addition, $ f_{bw}^h $ denotes the $ h $-th bandwidth in this filter. According to the equivalent rectangular bandwidth (ERB) \cite{glasberg1990derivation}, the bandwidth $ f_{bw} $ can be defined as a function of the center frequency $ f_c $ as follows:
\begin{equation}
\label{eq10}
\setlength{\abovedisplayskip}{6pt}
\setlength{\belowdisplayskip}{6pt}
f_{bw}\cong0.1079f_c+24.7.
\end{equation}
To more flexible optimization of the bandwidth, we add a learnable parameter $ \gamma $ into the \eqref{eq10}, and can be written as follows:
\begin{equation}
\label{eq11}
\setlength{\abovedisplayskip}{6pt}
\setlength{\belowdisplayskip}{6pt}
f_{bw}\cong\left(0.1079f_c+24.7\right)/\gamma,
\end{equation}
where $ \gamma $ is a learnable parameter and the initial value is 1. In addition, we set $ \gamma=max\left(\gamma,1\right) $. For the first harmonic filter, the center frequency $ f_c $ at the $ k $-th frequency bin can be computed with \eqref{eq5}. To satisfy the Nyquist criterion, for the first harmonic, the maximum frequency value is set as $ f_{max}=f_s/2H $, where $ f_s $ is the sampling rate and $ H $ is the number of harmonic. For any harmonic $ h > 1 $, the center frequency is scaled by the harmonic: $ h\cdot{f_c} $. Finally, the bandwidth of the $ h $-th harmonic can be defined as follows:

\begin{equation}
\label{eq13}
\setlength{\abovedisplayskip}{6pt}
\setlength{\belowdisplayskip}{6pt}
f_{bw}^h\cong\left(0.1079{h}\cdot {f_c}+24.7\right)/\gamma.
\end{equation}
Therefore, the designed harmonic filter bank can be defined as a combination of the above $ H $ harmonic filters in \eqref{eq9} as follows:
\begin{equation}
\label{eq14}
\setlength{\abovedisplayskip}{6pt}
\setlength{\belowdisplayskip}{6pt}
\nabla_h\left(f_c\right)|h=1,\cdots,H,f_c\in\left\{f_c^{\left(1\right)},\cdots,f_c^{\left(F\right)}\right\}.
\end{equation}

\sloppy
Therefore, we obtain a 3-dimensional harmonic tensor with the dimensionality $ \left[H,F,T\right] $, where each of them refers to the harmonic, frequency and time, respectively. This 3-dimensional harmonic tensor allows us to efficiently exploit locality information in the harmonic, frequency and time domains by utilizing the two-dimensional convolutional network. Furthermore, considering the existing energy below the "base" harmonic, i.e., $ h=1 $, one half harmonic ($ h=0.5 $) below the fundamental harmonic is added to further capture the fine-grained harmonic relationships at low-frequency bandwidths.
 
\sloppy
Furthermore according to \eqref{eq13}, we can note that the frequency bandwidth is dynamically variable and flexible due to the changing of the center frequency $ f_c $. In other words, as the center frequency $ f_c $ goes higher, the bandwidth $ f_{bw} $ goes wider, which results in a higher frequency resolution in the low-frequency band and a lower frequency resolution in the high-frequency band. This dynamic property facilitates the pitch modeling and fast-changing harmonic tracking. In addition, since we learn parameters of this harmonic filter in each sub-harmonic, these harmonic filter parameters can be efficiently optimized during the training process. Moreover, we limit the minimum value of the bandwidth parameter $ \gamma $ to 1, which can ensure that the maximal cutoff frequency satisfies the Nyquist criterion.

\sloppy
It is worth noting that the Constant-Q Transform (CQT) \cite{brown1992efficient} representation is also used as input of the discriminator, contributing to a dynamic frequency resolution for different frequency bands in \cite{gu2024multi}. However, our harmonic filter and CQT representation are considerably different. Specifically, since the center frequency at the $ k $-th frequency bin is measured by using \eqref{eq5}, harmonics $ h\cdot{f_c^k} $ can be learned only when $ h=2^d $, which results in the difficulty of odd harmonics modeling. Instead, our harmonic filter conveniently aligns the harmonically related content across the first dimension. Therefore, the $ k $-th frequency bin in the $ h $-th harmonic has the frequency $ {h\cdot{f_{min}}\cdot2^{k/B}} $, which is exactly the $ h $ multiple of the $ k $-th frequency bin in the first harmonic. This way of the harmonic aligning provides a complete harmonic representation, which can improve the ability of harmonic tracking and is further beneficial for high-quality speech generation. In addition, we use STFT operator with the fixed window length, which avoids the temporally asynchronous problem in CQT representation. Finally, instead of using the fixed bandwidth at each frequency bin in CQT,  we use the learnable bandwidth parameter $ \gamma $ to exploit the ideal bandwidth at each frequency bin in a data-driven manner.

\begin{figure}[t]
\centering
\includegraphics[width=0.44\textwidth]{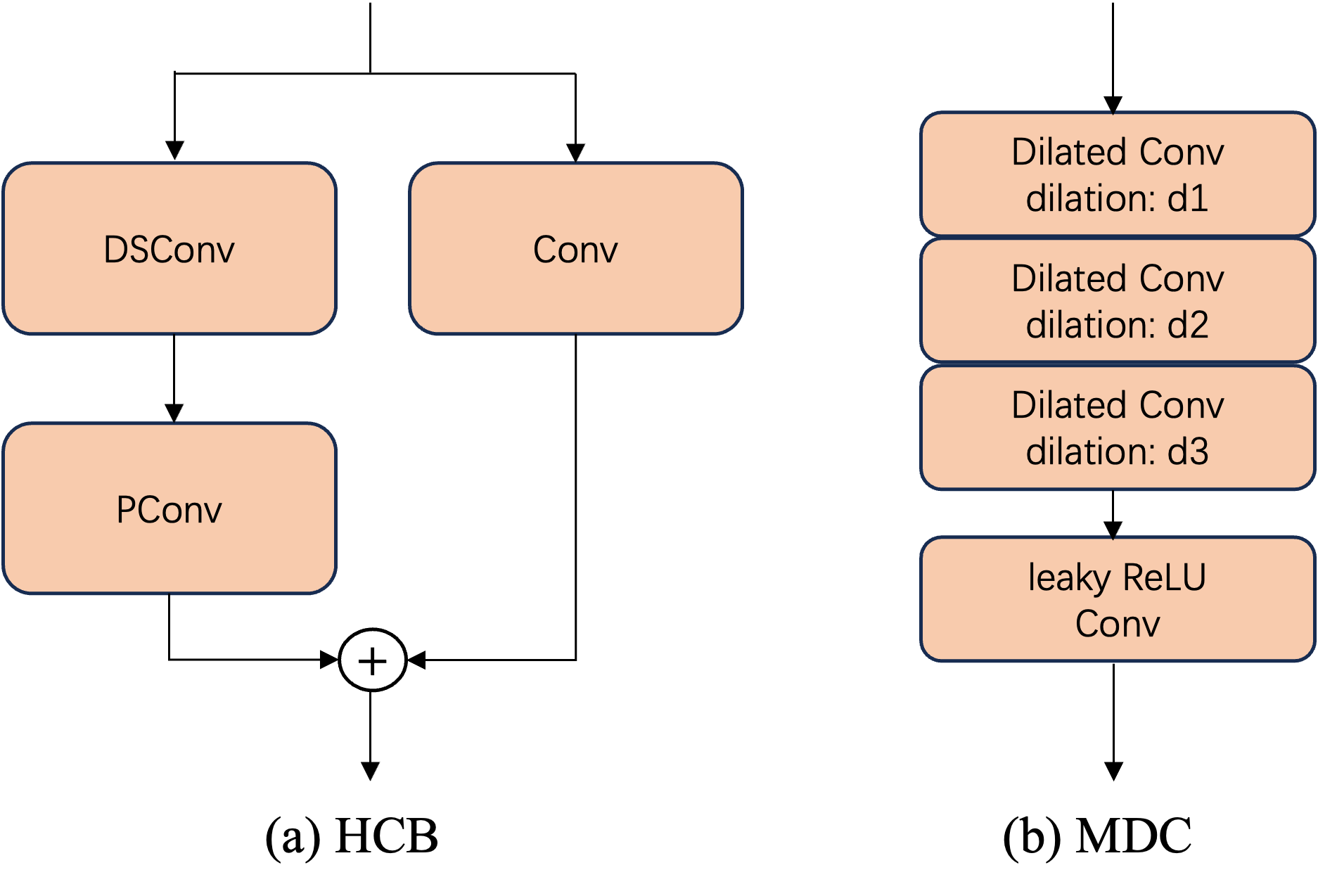}
\caption{The detailed architectures of the hybrid convolution block (HCB) and multi-scale dilated convolution (MDC). DSConv denotes the depthwise separable convolution and PConv denotes the pointwise convolution.}
\label{stru}
\end{figure}

\begin{table*}
\renewcommand{\arraystretch}{1.15}
\centering
\caption{\label{tab1}
	The experiment results of HiFiGAN and iSTFTNET vocoders on speech and singing datasets in terms of in-domain and out-of-domain samples. "S" and "C" denote MS-STFT and MS-SB-CQT discriminators, respectively. "H" denotes the proposed UnivHD discriminator. For each vocoder, the best results of different discriminators are listed in bold.}
\resizebox{1.0\textwidth}{!}{
\setlength{\tabcolsep}{1.5mm}{
\begin{tabular}{*{17}{c}}
  \toprule
  \multirow{3}*{{\bf Method}} & \multicolumn{8}{c}{{\bf speech dataset}} & \multicolumn{8}{c}{{\bf singing dataset}}\\  
  \cmidrule(lr){2-9}\cmidrule(lr){10-17}
  & \multicolumn{2}{c}{{\bf PESQ $\uparrow$}} & \multicolumn{2}{c}{{\bf MCD $\downarrow$}} & \multicolumn{2}{c}{{\bf F0RMSE $\downarrow$}} & \multicolumn{2}{c}{{\bf MOS $\uparrow$}} & \multicolumn{2}{c}{{\bf PESQ $\uparrow$}} & \multicolumn{2}{c}{{\bf MCD $\downarrow$}} & \multicolumn{2}{c}{{\bf F0RMSE $\downarrow$}} & \multicolumn{2}{c}{{\bf MOS $\uparrow$}} \\
  \cmidrule(lr){2-3}\cmidrule(lr){4-5}\cmidrule(lr){6-7}\cmidrule(lr){8-9}\cmidrule(lr){10-11}\cmidrule(lr){12-13}\cmidrule(lr){14-15}\cmidrule(lr){16-17}
  & {{\bf ID}} & {{\bf OD}} & {{\bf ID}} & {{\bf OD}} & {{\bf ID}} & {{\bf OD}} & {{\bf ID}} & {{\bf OD}} & {{\bf ID}} & {{\bf OD}} & {{\bf ID}} & {{\bf OD}} & {{\bf ID}} & {{\bf OD}} & {{\bf ID}} & {{\bf OD}}\\
  \midrule
  {Ground Truth }& 4.50 & 4.50 & 0.00 & 0.00 & - & - & 4.59$\pm$0.12 & 4.51$\pm$0.11 & 4.50 & 4.50 & 0.00 & 0.00 & - & - & 4.76$\pm$0.08 & 4.62$\pm$0.10 \\
   \midrule
  {HiFiGAN}& 3.00 & 2.81 & 2.99 & 1.75 & 46.25 & 53.43 & 3.85$\pm$0.13 & 3.81$\pm$0.15 & 2.96 & 2.66 & 2.45 & 2.69 & 30.32 & 50.96 & 3.37$\pm$0.09 & 3.48$\pm$0.07 \\
  {+S}& 3.02 & 2.82 & 2.95 & 1.76 & 41.06 & 50.23 & 3.91$\pm$0.14 & 3.77$\pm$0.13 & 3.05 & 2.70 & 2.36 & 2.68 & 30.28 & 46.08 & 3.56$\pm$0.06 & 3.66$\pm$0.09 \\
  {+C}& 3.06 & 2.82 & 2.92 & 1.70 & 39.46 & 44.77 & 3.98$\pm$0.13 & 3.83$\pm$0.16 & 3.10 & 2.74 & 2.29 & 2.67 & 29.98 & 44.96 & 3.67$\pm$0.08 & 3.75$\pm$0.06 \\
  {+H}& {\bf 3.13} & {\bf 2.91} & {\bf 2.88} & {\bf 1.69} & {\bf 37.85} & {\bf 42.88} & {\bf 4.05$\pm$0.13} & {\bf 3.97$\pm$0.14} & {\bf 3.19} & {\bf 2.85} & {\bf 2.27} & {\bf 2.54} & {\bf 28.94} & {\bf 43.04} & {\bf 3.78$\pm$0.06} & {\bf3.86$\pm$0.09} \\
 \midrule
  {iSTFTNET}& 2.95 & 2.81 & 2.96 & 1.68 & 43.41 & 52.91 & 3.91$\pm$0.14 & 3.85$\pm$0.12 & 3.06 & 2.81 & 2.16 & 2.39 & 28.96 & 49.03 & 3.51$\pm$0.08 & 3.62$\pm$0.09 \\
  {+S}& 2.99 & 2.79 & 2.96 & 1.72 & 40.46 & 49.27 & 3.95$\pm$0.16 & 3.80$\pm$0.17 & 3.05 & 2.82 & 2.20 & 2.34 & 27.79 & 43.95 & 3.62$\pm$0.11 & 3.77$\pm$0.08 \\
  {+C}& 3.01 & 2.84 & 2.94 & 1.63 & 41.03 & 46.71 & 4.01$\pm$0.14 & 3.91$\pm$0.15 & 3.09 & 2.87 & 2.12 & 2.31 & 27.40 & 40.20 & 3.75$\pm$0.07 & 3.84$\pm$0.11 \\
  {+H}& {\bf 3.07} & {\bf 2.88} & {\bf 2.92} & {\bf 1.56} & {\bf 40.03} & {\bf 45.89} & {\bf 4.03$\pm$0.13} & {\bf 3.99$\pm$0.16} & {\bf 3.14} & {\bf 2.90} & {\bf 2.09} & {\bf 2.25} & {\bf 26.32} & {\bf 38.79} & {\bf 3.84$\pm$0.11} & {\bf 3.91$\pm$0.08} \\
  \bottomrule
\end{tabular}}}
\end{table*}

\subsection{Model Architecture}
\label{str}
\sloppy
The designed harmonic tensor has three dimensions (harmonic, frequency and time). As a result, we utilize two-dimensional convolutional layers to efficiently learn locality in terms of the frequency and time domains. In terms of the harmonic domain, we employ the combination of depthwise separable convolution and normal convolution layers to simultaneously learn the intra-harmonic and inter-harmonic information in speech.

\sloppy
The overview of the discriminator network architecture is illustrated in Fig.~\ref{mhd}. After the harmonic filter bank, we obtain the real-valued representation, i.e., harmonic tensor. Next, we take this harmonic tensor as input of the hybrid convolution block (HCB) to get the latent representation. Then, the generated latent representation is sequentially fed into three multi-scale dilated convolution (MDC) blocks and one convolution layer. Finally, the outputs of each MDC block and the final convolution layer are used to compute the feature matching loss and discriminator loss, respectively. 

\sloppy
The detailed network architectures of the hybrid convolution block (HCB) and multi-scale dilated convolution (MDC) are illustrated in Fig.~\ref{stru}. For HCB, the harmonic is considered as channel axis and thus we employ the depthwise separable convolution (DSConv) and pointwise convolution (PConv) to exploit the intra-harmonic structure of speech. Additionally, we utilize the normal convolution layer to capture the relationship among harmonics. Next, we add their outputs to generate the latent representation. Furthermore, each MDC contains three dilated convolution layers and a normal convolution layer with a leaky ReLU activation function. To cover diverse receptive fields, we utilize different dilation rates in each MDC. It is worth noting that these convolution layers are all two-dimensional.

\begin{figure*}[t]
\centering
\includegraphics[width=0.85\textwidth]{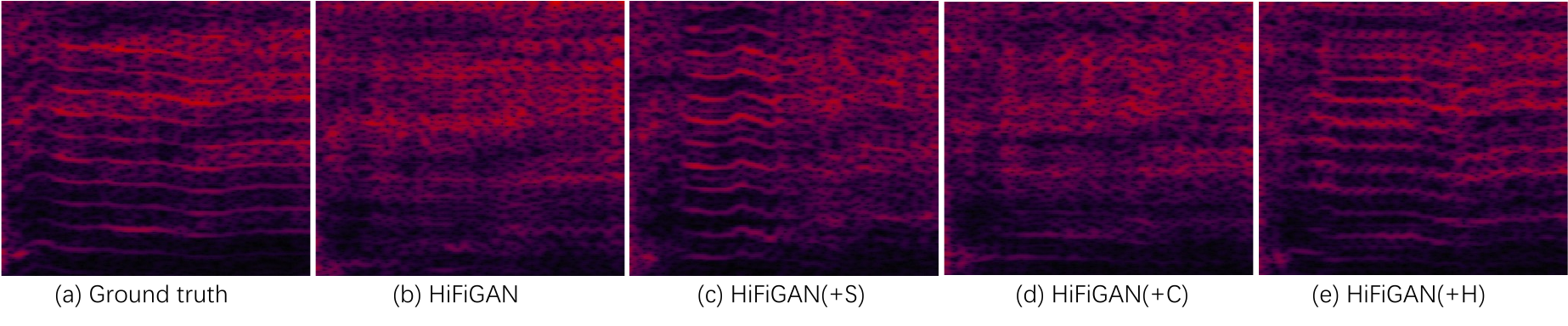}
\caption{The spectrogram visualization of an out-of-domain singing voice sample in terms of different time-frequency based discriminators for HiFiGAN. The zoomed-in high-frequency views are presented, respectively.}
\label{vis}
\end{figure*}

\section{Experiments}
\label{exp}

\subsection{Datasets}
\label{data}
\sloppy
For training, we use the speech and singing voices as our training datasets, respectively. For the speech dataset, we utilize the train-clean-100 dataset from LibriTTS \cite{koizumi2023libritts}. For the singing dataset, we use the OpenSinger \cite{huang2021multi} dataset. For testing, we construct two datasets for speech and singing voices, named in-domain (ID) and out-of-domain (OD) datasets, respectively. Specifically for speech, we randomly select 500 utterances from LibriTTS forming ID dataset and the remaining samples are used for training. For evaluating unseen speakers, 500 utterances are also chosen from VCTK \cite{yamagishi2019cstr} to form the OD dataset. For the singing voices, we also randomly sample 500 utterances from OpenSinger as ID dataset, which contains seen singers. For the OD dataset, 500 samples from M4Singer \cite{zhang2022m4singer} and Opencpop \cite{wang2022opencpop} datasets are randomly selected to evaluate unseen singers.

\subsection{Training Setups and Evaluations}
\label{tb}
\sloppy
\noindent{\bf Implementation Details.} The output channel of DSConv in HCB is the same as the number of harmonic $ H $. The output channels of PConv and normal convolution layers in HCB are both 32. Their kernel sizes are all set as (7, 7). In each MDC module, the dilation factors of three dilated convolution are 1, 2 and 4, respectively. The kernel size is (5, 5) for all convolution layers in MDC, and strides in dilated and normal convolution layers are (1, 1) and (2, 1), respectively. The output channels of each MDC are both set as 32. For the final convolution layer, the kernel size along the frequency axis is equal to the output feature dimension of MDC blocks. Furthermore, the number of bins per octave is set as 24 and $ H $ is set as 10, which leads to 124 frequency bins within each harmonic range. The parameter of UnivHD is 0.31 M. All training samples are resampled at 24k Hz and all models are trained up to 1.5 million steps.

\sloppy
\noindent{\bf Baselines and Evaluations.} We use two time-frequency based discriminators as baselines, including Multi-Scale STFT (MS-STFT)\cite{defossez2022high} and Multi-Scale Sub-Band CQT (MS-SB-CQT) discriminator \cite{gu2024multi}. The configurations of these discriminators are the same as the official versions. In addition, we use the HiFiGAN\footnote{\url{https://github.com/jik876/hifi-gan}} \cite{kong2020hifi} and iSTFTNET \footnote{\url{https://github.com/rishikksh20/iSTFTNet-pytorch}} \cite{kaneko2022istftnet} as vocoders and maintain the same training configurations with the original versions other than adding additional time-frequency based discriminator. For evaluations, we employ the mel-cepstral distortion (MCD)\footnote{\url{https://github.com/chenqi008/pymcd}} \cite{kubichek1993mel} with dynamic time warping and the Perceptual Evaluation of Speech Quality (PESQ)\footnote{\url{https://github.com/ludlows/python-pesq}} \cite{rix2001perceptual} to estimate the spectrogram reconstruction and F0 Root Mean Square Error (F0RMSE)\footnote{\url{https://github.com/gemelo-ai/vocos}} for pitch error evaluating. For the subjective metric, the 5-point Mean Opinion Score (MOS) is utilized to evaluate the speech quality. Score 1 denotes poor speech and 5 denotes excellent speech. Specifically, 15 utterances are randomly selected from each test dataset for MOS test and a total of ten people participate. Participants are required to evaluate each utterance once.

\subsection{Results}
\label{resu}

\subsubsection{Discriminator Performance}
\sloppy
We first estimate the performance of the proposed UnivHD discriminator based on the HiFiGAN and iSTFTNET vocoders, as illustrated in Table~\ref{tab1}. All time-frequency based discriminators achieve the comparable or superior performance compared to the original versions of vocoders, especially on the singing datasets. This finding indicates that these time-frequency based discriminators contribute to the improvement of speech quality and intelligibility. In addition, MS-SB-CQT discriminator and the proposed UnivHD surpass the MS-STFT discriminator on the singing datasets in terms of all metrics. The primary reason is that the former has the dynamic resolution in frequency domain and MS-STFT has a fixed resolution across all frequency bands instead. This finding highlights the importance of dynamic frequency resolution in the discriminator, especially for expressive singing voices. Furthermore, our naive version of UnivHD also performs better than the original vocoder versions and achieves the competitive performance compared to MS-SB-CQT. When applying the proposed UnivHD, the significant superiority is shown compared to all baseline discriminators, especially in terms of boosting in MOS, which indicates the importance of the complete harmonic space and the effectiveness of the learnable harmonic filter. 

\begin{figure}[t]
\centering
\includegraphics[width=0.486\textwidth]{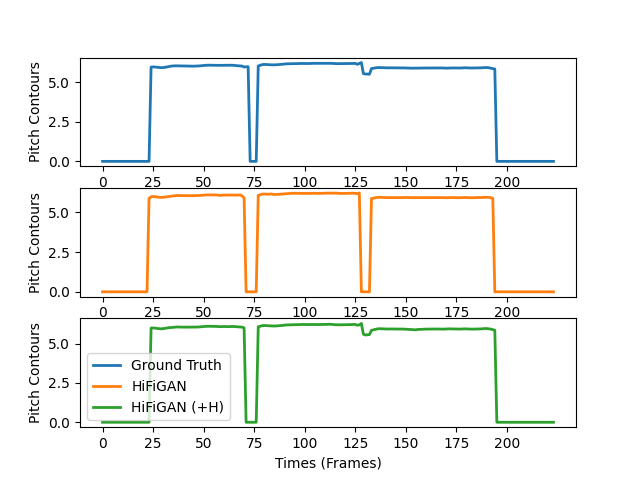}
\caption{The pitch contours without the normalization of an out-of-domain singing voice sample.}
\label{ff}
\end{figure}

\sloppy
We also select a singing sample in the out-of-domain singing dataset for the spectrogram visualization and the results of high-frequency views are presented in Fig.~\ref{vis}. Notably, aliasing artifacts are commonly observed in GAN-based vocoders, like HiFiGAN in Fig.~\ref{vis}b. Furthermore, adding the MS-STFT discriminator can provide an accurate reconstruction of frequency band. However, as shown in Fig.~\ref{vis}c, its distribution of harmonic is inconsistent compared to ground truth due to the fixed frequency resolution. MS-SB-CQT discriminator cannot capture odd harmonics, which also suffers from aliasing artifacts in high-frequency band (Fig.~\ref{vis}d). Conversely, HiFi-GAN with the UnivHD (Fig.~\ref{vis}e) can recover these harmonics and track them, which makes an improvement of the original HiFiGAN. Additionally, Fig.~\ref{ff} shows the pitch contours of this sample. For the original HiFiGAN, the pitch contour mismatches the pitch of ground truth at the 125-th frame. In contrast, UnivHD rectifies the pitch contour, which is located at the ground truth position as expected. This finding highlights that the proposed harmonic filter can also correctly focus on the lower part of the spectrogram. 

Furthermore, we combine the MS-STFT discriminator and the proposed UnivHD, i.e., HiFiGAN (+S+H). The objective results on the speech dataset are illustrated in Table~\ref{tab2}. Notably, combining the proposed harmonic filter and STFT yields performance gains. This finding confirms that the complementary information in different discriminators is jointly optimized in the training process, which further facilitates the improvement of the generated speech quality. Additionally, HiFiGAN (+S+H) achieves better performance compared to HiFiGAN (+S+C), which also verifies the superior performance of the proposed harmonic discriminator.

\begin{table}
\renewcommand{\arraystretch}{1.1}
\centering
\caption{\label{tab2}
	The objective experiment results of the combined discriminators on the speech dataset.} 
\resizebox{0.486\textwidth}{!}{
\setlength{\tabcolsep}{2.5mm}{
\begin{tabular}{*{5}{c}}
  \toprule
  \multirow{2}*{{\bf Method}} & \multicolumn{2}{c}{{\bf speech ID}} & \multicolumn{2}{c}{{\bf speech OD}} \\  
  \cmidrule(lr){2-3}\cmidrule(lr){4-5}
  & {{\bf PESQ $\uparrow$}} & {{\bf MCD $\downarrow$}} & {{\bf PESQ $\uparrow$}} & {{\bf MCD $\downarrow$}} \\
  \midrule
  {HiFiGAN (+H)}& 3.13 & 2.88 & 2.91 & 1.69 \\ 
  {HiFiGAN (+S+C)}& 3.09 & 2.88 & 2.86 & 1.67 \\ 
  {HiFiGAN (+S+H)}& {\bf 3.15} & {\bf 2.84} & {\bf 2.94} & {\bf 1.65} \\ 
  \bottomrule
\end{tabular}}}
\end{table}

\subsubsection{Ablation Study}

\begin{figure}[t]
\centering
\includegraphics[width=0.486\textwidth]{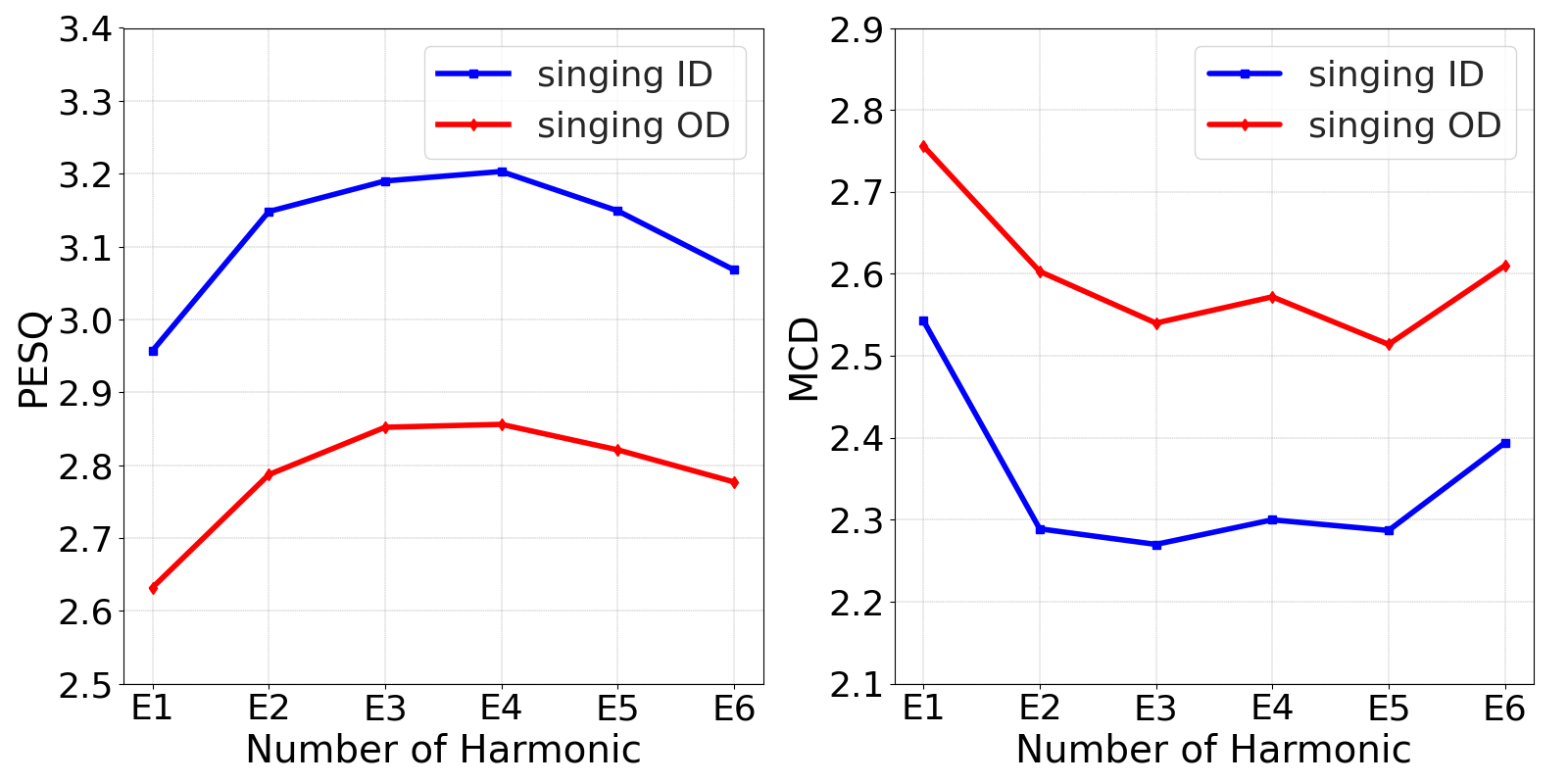}
\caption{The results of ablation experiments in terms of different numbers of harmonic filters on the singing dataset.}
\label{dis}
\end{figure}

\sloppy
\noindent{\bf Harmonic Number Ablation.} We explore the influence of different numbers of harmonic $ H $ in the proposed UnivHD and the ablation experiment results based on HiFiGAN are reported in Fig.~\ref{dis}. Specifically, E1-E6 refer to different numbers of harmonic $ H $ that are set as 2, 8, 10, 12, 15 and 20, respectively. As shown in Fig.~\ref{dis}, different numbers of harmonic can achieve the approximate performances except for smaller or larger values, which indicates the robustness to this parameter. Additionally, the smaller $ H $, e.g., 2, will obtain a coarse segmentation in the frequency band, leading to the degraded quality. The larger $ H $, e.g., 20, also remains the limitation of fundamental frequency modeling. For example, there is a speech sample at 24k Hz sampling rate and the fundamental frequency value is higher than 600 Hz. Therefore, this harmonic discriminator cannot learn the fundamental frequency within the first harmonic for $ H>20 $, which significantly degrades the performance.

\sloppy
\noindent{\bf Discriminator Architecture Ablation.} We conduct ablation experiments of the model architecture based on the singing dataset. As illustrated in Table~\ref{tab3}, omitting each proposed component can result in performance decline. Notably, removing DSConv leads to a significant increase on the magnitude-related MCD and pitch-related F0RMSE metrics, which validates the effectiveness of DSConv in capturing intra-harmonic relationships. Removing the normal convolution results in a significant decline on PESQ metric, which indicates the importance of the global harmonic for the improvement of speech quality. Additionally, omitting the half-harmonic representation leads to a slight performance decline. The primary reason is that the half-harmonic representation captures the fine-grained harmonic relationships below the fundamental harmonic ($ H=1 $). The energy below the first harmonic can be effectively learned with the half representation, which contributes to the generation of the accurate pitch details and high-quality speech.

\begin{table}
\renewcommand{\arraystretch}{1.1}
\centering
\caption{\label{tab3}
	The objective results of architecture ablation experiments on the singing out-of-domain dataset.}
\resizebox{0.486\textwidth}{!}{
\setlength{\tabcolsep}{3.7mm}{
\begin{tabular}{*{5}{c}}
  \toprule
  {{\bf Method}} & {{\bf PESQ $\uparrow$}} & {{\bf MCD $\downarrow$}} & {{\bf F0RMSE $\downarrow$}}\\
  \midrule
  {HiFiGAN (+H)}& {\bf 2.85} & {\bf 2.54} & {\bf 43.04} \\ 
  {w/o. DSConv}& 2.82 & 2.75 & 45.72 \\ 
  {w/o. normal conv}& 2.71 & 2.65 & 44.55 \\ 
  {w/o. dilate conv}& 2.77 & 2.59 & 43.97 \\ 
  {w/o. half-harmonic}& 2.83 & 2.55 & 45.13 \\ 
  \bottomrule
\end{tabular}}}
\end{table}

\section{Conclusion}
\label{con}
\sloppy
In this paper, we design a learnable harmonic filter and integrate it into the discriminator of vocoder for high-quality speech generation. In our method, we extend the triangular band-pass filter to the harmonic filter that owns the learnable parameters and dynamic frequency resolution. Moreover, the center frequencies of this harmonic filter are scaled by the harmonic orders, which forms the complete harmonic space and further improves the quality of generated speech. Additionally, we add a half-harmonic representation to learn the energy at low-frequency band. Experimental results also demonstrate the effectiveness of the proposed harmonic discriminator and the half-harmonic representation.



\bibliographystyle{IEEEtran}
\bibliography{mybib}

@inproceedings{lee-etal-2022-direct,
    title = {Direct Speech-to-Speech Translation With Discrete Units},
    author = {Lee, Ann and Chen, Peng-Jen and Wang, Changhan and Gu, Jiatao and Popuri, Sravya and Ma, Xutai and Polyak, Adam and Adi, Yossi and He, Qing and Tang, Yun and Pino, Juan and Hsu, Wei-Ning},
    booktitle = {Proceedings of the Association for Computational Linguistics (ACL)},
    year = {2022},
    pages = {3327--3339}
}

@inproceedings{ren2019fastspeech,
  title={Fastspeech: Fast, robust and controllable text to speech},
  author={Ren, Yi and Ruan, Yangjun and Tan, Xu and Qin, Tao and Zhao, Sheng and Zhao, Zhou and Liu, Tie-Yan},
  booktitle={Advances in Neural Information Processing Systems (NeurIPS)},
  volume={32},
  pages={3171--3180},
  year={2019}
}

@inproceedings{bai20223,
  title={A3T: Alignment-Aware Acoustic and Text Pretraining for Speech Synthesis and Editing},
  author={Bai, He and Zheng, Renjie and Chen, Junkun and Ma, Mingbo and Li, Xintong and Huang, Liang},
  booktitle={International Conference on Machine Learning (ICML)},
  pages={1399--1411},
  year={2022}
}

@inproceedings{jiang2024mega,
  title={Mega-TTS 2: Boosting Prompting Mechanisms for Zero-Shot Speech Synthesis},
  author={Jiang, Ziyue and Liu, Jinglin and Ren, Yi and He, Jinzheng and Ye, Zhenhui and Ji, Shengpeng and Yang, Qian and Zhang, Chen and Wei, Pengfei and Wang, Chunfeng and Yin, Xiang and Ma, Zejun and Zhao, Zhou},
  booktitle={International Conference on Learning Representations (ICLR)},
  year={2024}
}

@inproceedings{huang2023make,
  title={Make-an-audio: Text-to-audio generation with prompt-enhanced diffusion models},
  author={Huang, Rongjie and Huang, Jiawei and Yang, Dongchao and Ren, Yi and Liu, Luping and Li, Mingze and Ye, Zhenhui and Liu, Jinglin and Yin, Xiang and Zhao, Zhou},
  booktitle={International Conference on Machine Learning (ICML)},
  pages={13916--13932},
  year={2023}
}

@article{liu2024audioldm,
  title={Audioldm 2: Learning holistic audio generation with self-supervised pretraining},
  author={Liu, Haohe and Yuan, Yi and Liu, Xubo and Mei, Xinhao and Kong, Qiuqiang and Tian, Qiao and Wang, Yuping and Wang, Wenwu and Wang, Yuxuan and Plumbley, Mark D},
  journal={IEEE/ACM Transactions on Audio, Speech, and Language Processing (TASLP)},
  volume={32},
  pages={2871--2883},
  year={2024}
}

@inproceedings{Lee2025period,
  title={PeriodWave: Multi-Period Flow Matching for High-Fidelity Waveform Generation},
  author={Sang-Hoon, Lee and Ha-Yeong, Choi and Seong-Whan, Lee},
  booktitle={International Conference on Learning Representations (ICLR)},
  year={2025}
}

@inproceedings{kumar2019melgan,
  title={Melgan: Generative adversarial networks for conditional waveform synthesis},
  author={Kumar, Kundan and Kumar, Rithesh and De Boissiere, Thibault and Gestin, Lucas and Teoh, Wei Zhen and Sotelo, Jose and De Brebisson, Alexandre and Bengio, Yoshua and Courville, Aaron C},
  booktitle={Advances in Neural Information Processing Systems (NeurIPS)},
  volume={32},
  pages={14910--14921},
  year={2019}
}

@inproceedings{kong2020hifi,
  title={Hifi-gan: Generative adversarial networks for efficient and high fidelity speech synthesis},
  author={Kong, Jungil and Kim, Jaehyeon and Bae, Jaekyoung},
  booktitle={Advances in Neural Information Processing Systems (NeurIPS)},
  volume={33},
  pages={17022--17033},
  year={2020}
}

@misc{tian2020tfgan,
      title={TFGAN: Time and Frequency Domain Based Generative Adversarial Network for High-fidelity Speech Synthesis},
      author={Qiao Tian and Yi Chen and Zewang Zhang and Heng Lu and Linghui Chen and Lei Xie and Shan Liu},
      year={2020},
      eprint={2011.12206},
      archivePrefix={arXiv},
      url={https://arxiv.org/abs/2011.12206},
}

@inproceedings{Kazuki2021hwg,
  title={Harmonic WaveGAN: GAN-Based Speech Waveform Generation Model with
Harmonic Structure Discriminator},
  author={Mizuta, Kazuki and Koriyama, Tomoki and Saruwatari, Hiroshi},
  booktitle={Interspeech},
  pages={2192--2196},
  year={2021}
}

@inproceedings{jang2021univnet,
  title={Univnet: A neural vocoder with multi-resolution spectrogram discriminators for high-fidelity waveform generation},
  author={Jang, Won and Lim, Dan and Yoon, Jaesam and Kim, Bongwan and Kim, Juntae},
  booktitle={Interspeech},
  pages={2207--2211},
  year={2021}
}

@inproceedings{leebigvgan,
  title={BigVGAN: A Universal Neural Vocoder with Large-Scale Training},
  author={Lee, Sang-gil and Ping, Wei and Ginsburg, Boris and Catanzaro, Bryan and Yoon, Sungroh},
  booktitle={International Conference on Learning Representations (ICLR)},
  year={2023}
}

@inproceedings{song2023dspgan,
  title={Dspgan: a gan-based universal vocoder for high-fidelity tts by time-frequency domain supervision from dsp},
  author={Song, Kun and Zhang, Yongmao and Lei, Yi and Cong, Jian and Li, Hanzhao and Xie, Lei and He, Gang and Bai, Jinfeng},
  booktitle={IEEE International Conference on Acoustics, Speech and Signal Processing (ICASSP)},
  pages={1--5},
  year={2023}
}

@inproceedings{li2023snakegan,
  title={SnakeGAN: A Universal Vocoder Leveraging DDSP Prior Knowledge and Periodic Inductive Bias},
  author={Li, Sipan and Liu, Songxiang and Zhang, Luwen and Li, Xiang and Bian, Yanyao and Weng, Chao and Wu, Zhiyong and Meng, Helen},
  booktitle={IEEE International Conference on Multimedia and Expo (ICME)},
  pages={1703--1708},
  year={2023}
}

@inproceedings{bak2023avocodo,
  title={Avocodo: Generative adversarial network for artifact-free vocoder},
  author={Bak, Taejun and Lee, Junmo and Bae, Hanbin and Yang, Jinhyeok and Bae, Jae-Sung and Joo, Young-Sun},
  booktitle={Proceedings of the AAAI Conference on Artificial Intelligence (AAAI)},
  volume={37},
  pages={12562--12570},
  year={2023}
}

@inproceedings{kaneko2022istftnet,
  title={iSTFTNet: Fast and lightweight mel-spectrogram vocoder incorporating inverse short-time Fourier transform},
  author={Kaneko, Takuhiro and Tanaka, Kou and Kameoka, Hirokazu and Seki, Shogo},
  booktitle={IEEE International Conference on Acoustics, Speech and Signal Processing (ICASSP)},
  pages={6207--6211},
  year={2022}
}

@inproceedings{kaneko2023istftnet2,
  title={iSTFTNet2: Faster and more lightweight iSTFT-based neural vocoder using 1D-2D CNN},
  author={Kaneko, Takuhiro and Kameoka, Hirokazu and Tanaka, Kou and Seki, Shogo},
  booktitle={Interspeech},
  pages={4369--4373},
  year={2023}
}

@inproceedings{siuzdak2023vocos,
  title={Vocos: Closing the gap between time-domain and Fourier-based neural vocoders for high-quality audio synthesis},
  author={Siuzdak, Hubert},
  booktitle={International Conference on Learning Representations (ICLR)},
  year={2024}
}

@article{ai2020neural,
  title={A neural vocoder with hierarchical generation of amplitude and phase spectra for statistical parametric speech synthesis},
  author={Ai, Yang and Ling, Zhen-Hua},
  journal={IEEE/ACM Transactions on Audio, Speech, and Language Processing (TASLP)},
  volume={28},
  pages={839--851},
  year={2020}
}

@article{ai2023apnet,
  title={APNet: An all-frame-level neural vocoder incorporating direct prediction of amplitude and phase spectra},
  author={Ai, Yang and Ling, Zhen-Hua},
  journal={IEEE/ACM Transactions on Audio, Speech, and Language Processing (TASLP)},
  volume={31},
  pages={2145--2157},
  year={2023}
}

@inproceedings{du2023apnet2,
  title={APNet2: High-Quality and High-Efficiency Neural Vocoder with Direct Prediction of Amplitude and Phase Spectra},
  author={Du, Hui-Peng and Lu, Ye-Xin and Ai, Yang and Ling, Zhen-Hua},
  booktitle={National Conference on Man-Machine Speech Communication (NCMMSC)},
  pages={66--80},
  year={2023}
}

@article{nguyen1994near,
  title={Near-perfect-reconstruction pseudo-QMF banks},
  author={Nguyen, Truong Q},
  journal={IEEE Transactions on Signal Processing},
  volume={42},
  number={1},
  pages={65--76},
  year={1994}
}

@article{wang2003modified,
  title={Modified discrete cosine transform: Its implications for audio coding and error concealment},
  author={Wang, Ye and Vilermo, Mikka},
  journal={Journal of the Audio Engineering Society (AES)},
  volume={51},
  number={1/2},
  pages={52--61},
  year={2003}
}

@article{defossez2022high,
  title={High fidelity neural audio compression},
  author={D{\'e}fossez, Alexandre and Copet, Jade and Synnaeve, Gabriel and Adi, Yossi},
  journal={Transactions on Machine Learning Research (TMLR)},
  year={2023}
}

@inproceedings{wang2022opencpop,
  title={Opencpop: A high-quality open source chinese popular song corpus for singing voice synthesis},
  author={Wang, Yu and Wang, Xinsheng and Zhu, Pengcheng and Wu, Jie and Li, Hanzhao and Xue, Heyang and Zhang, Yongmao and Xie, Lei and Bi, Mengxiao},
  booktitle={Interspeech},
  pages={4242--4246},
  year={2022}
}

@inproceedings{pariente2020filterbank,
  title={Filterbank design for end-to-end speech separation},
  author={Pariente, Manuel and Cornell, Samuele and Deleforge, Antoine and Vincent, Emmanuel},
  booktitle={IEEE International Conference on Acoustics, Speech and Signal Processing (ICASSP)},
  pages={6364--6368},
  year={2020}
}

@inproceedings{won2020data,
    title={Data-driven harmonic filters for audio representation learning},
    author={Won, Minz and Chun, Sanghyuk and Nieto, Oriol and Serrc, Xavier},
    booktitle={IEEE International Conference on Acoustics, Speech and Signal Processing (ICASSP)},
    pages={536--540},
    year={2020}
}

@inproceedings{ho2024neural,
  title={What do neural networks listen to? Exploring the crucial bands in Speech Enhancement using Sinc-convolution},
  author={Ho, Kuan-Hsun and Hung, Jeih-weih and Chen, Berlin},
  booktitle={IEEE International Conference on Acoustics, Speech and Signal Processing (ICASSP)},
  pages={10406--10410},
  year={2024}
}

@article{glasberg1990derivation,
  title={Derivation of auditory filter shapes from notched-noise data},
  author={Glasberg, Brian R and Moore, Brian CJ},
  journal={Hearing Research},
  volume={47},
  pages={103--138},
  year={1990}
}

@inproceedings{gu2024multi,
  title={Multi-scale sub-band constant-q transform discriminator for high-fidelity vocoder},
  author={Gu, Yicheng and Zhang, Xueyao and Xue, Liumeng and Wu, Zhizheng},
  booktitle={IEEE International Conference on Acoustics, Speech and Signal Processing (ICASSP)},
  pages={10616--10620},
  year={2024}
}

@article{brown1992efficient,
  title={An efficient algorithm for the calculation of a constant Q transform},
  author={Brown, Judith C and Puckette, Miller S},
  journal={The Journal of the Acoustical Society of America (JASA)},
  volume={92},
  number={5},
  pages={2698--2701},
  year={1992}
}

@inproceedings{kubichek1993mel,
  title={Mel-cepstral distance measure for objective speech quality assessment},
  author={Kubichek, Robert},
  booktitle={Proceedings of IEEE Pacific Rim Conference on Communications, Computers and Signal Processing (PACRIM)},
  volume={1},
  pages={125--128},
  year={1993}
}

@inproceedings{schorkhuber2010constant,
  title={Constant-Q transform toolbox for music processing},
  author={Sch{\"o}rkhuber, Christian and Klapuri, Anssi},
  booktitle={Sound and Music Computing},
  pages={3--64},
  year={2010}
}

@inproceedings{rix2001perceptual,
  title={Perceptual evaluation of speech quality (PESQ)-a new method for speech quality assessment of telephone networks and codecs},
  author={Rix, Antony W and Beerends, John G and Hollier, Michael P and Hekstra, Andries P},
  booktitle={IEEE International Conference on Acoustics, Speech and Signal Processing (ICASSP)},
  volume={2},
  pages={749--752},
  year={2001}
}

@inproceedings{koizumi2023libritts,
  title={Libritts-r: A restored multi-speaker text-to-speech corpus},
  author={Koizumi, Yuma and Zen, Heiga and Karita, Shigeki and Ding, Yifan and Yatabe, Kohei and Morioka, Nobuyuki and Bacchiani, Michiel and Zhang, Yu and Han, Wei and Bapna, Ankur},
  booktitle={Interspeech},
  pages={5496--5500},
  year={2023}
}

@inproceedings{huang2021multi,
  title={Multi-singer: Fast multi-singer singing voice vocoder with a large-scale corpus},
  author={Huang, Rongjie and Chen, Feiyang and Ren, Yi and Liu, Jinglin and Cui, Chenye and Zhao, Zhou},
  booktitle={Proceedings of ACM International Conference on Multimedia (ACM MM)},
  pages={3945--3954},
  year={2021}
}

@article{yamagishi2019cstr,
  title={CSTR VCTK Corpus: English multi-speaker corpus for CSTR voice cloning toolkit (version 0.92)},
  author={Yamagishi, Junichi and Veaux, Christophe and MacDonald, Kirsten},
  journal={University of Edinburgh. The Centre for Speech Technology Research (CSTR)},
  pages={271--350},
  year={2019}
}

@article{zhang2022m4singer,
  title={M4singer: A multi-style, multi-singer and musical score provided mandarin singing corpus},
  author={Zhang, Lichao and Li, Ruiqi and Wang, Shoutong and Deng, Liqun and Liu, Jinglin and Ren, Yi and He, Jinzheng and Huang, Rongjie and Zhu, Jieming and Chen, Xiao and Zhao, Zhou},
  journal={Advances in Neural Information Processing Systems (NeurIPS)},
  volume={35},
  pages={6914--6926},
  year={2022}
}



\end{document}